\documentclass[12pt,preprint]{aastex}

\usepackage{epsfig}
\usepackage{graphicx}
\usepackage{amsmath}
\usepackage{amssymb}
\usepackage{appendix}

\def\be{\begin{equation}}
\def\ee{\end{equation}}
\def\ba{\begin{eqnarray}}
\def\ea{\end{eqnarray}}
\def\go{\mathrel{\raise.3ex\hbox{$>$}\mkern-14mu
             \lower0.6ex\hbox{$\sim$}}}
\def\lo{\mathrel{\raise.3ex\hbox{$<$}\mkern-14mu
             \lower0.6ex\hbox{$\sim$}}}

\begin{document}

\title{Formation of Super-Earths by Tidally-Forced Turbulence
}

\author{Cong Yu\altaffilmark{1}}
\altaffiltext{1}
{School of Physics and Astronomy, Sun Yat-Sen University, Guangzhou, 519082, P. R. China;
{\tt yucong@mail.sysu.edu.cn}
}


\begin{abstract}
The Kepler observations indicate that many exoplanets are super-Earths, which brings about a puzzle for the core-accretion scenario. Since observed super-Earths are in the range of critical mass, they would accrete gas efficiently and become gas giants. Theoretically, super-Earths are predicted to be rare in the core-accretion framework. To resolve this contradiction, we propose that the tidally-forced turbulent diffusion may affect the heat transport inside the planet. Thermal feedback induced by turbulent diffusion is investigated. We find that the tidally-forced turbulence would generate pseudo-adiabatic regions within radiative zones, which pushes the radiative-convective boundaries (RCBs) inwards. This would decrease the cooling luminosity and enhance the Kelvin-Helmholtz (KH) timescale. For a given lifetime of protoplanetary disks (PPDs), there exists a critical threshold for the turbulent diffusivity, $\nu_{\rm critical}$. If $\nu_{\rm turb}>\nu_{\rm critical} $, the KH timescale is longer than the disk lifetime and the planet would become a super-Earth rather than a gas giant. We find that even a small value of turbulent diffusion has influential effects on evolutions of super-Earths. $\nu_{\rm critical}$ increases with the core mass. We further ascertain that, within the minimum mass extrasolar nebula (MMEN), $\nu_{\rm critical}$ increases with the semi-major axis. This may explain the feature that super-Earths are common in inner PPD regions, while gas giants are common in the outer PPD regions. The predicted envelope mass fraction (EMF) is not fully consistent with observations. We discuss physical processes, such as late core assembly and mass loss mechanisms, that may be operating during super-Earth formation.
\end{abstract}


\section{Introduction}
One of the widely accepted mechanisms
of planet formation is the core-nucleated instability theory (Perri \& Cameron 1974; Haris 1978; Mizuno
et al. 1978; Stevenson 1982). According to this scenario, the massive gaseous atmosphere would
be accumulated in a runaway manner when the core mass reaches a critical value.
In static models, when heating balances cooling,
runaway accretion occurs when the planet is beyond the critical mass,
because the envelope fails to hold hydrostatic equilibrium. Rafikov (2006) found
a broad range of critical mass ($0.1 M_{\oplus} \le M_{\rm critical} \le 100 M_{\oplus}$)
due to various disk properties and planetesimal accretion rate.
However, in dynamic or quasi-static models, the thermal disequilibrium
rather than the hydrostatic disequilibrium plays the dominant role.
The runaway accretion occurs because the envelope becomes thermally unstable
as the cooling timescale becomes catastrophically shorter.
In this case, the runaway accretion is driven by
runaway cooling (Bodenheimer \& Pollack 1986; Lee et al. 2014; Piso \& Youdin 2014).
Three stages are involved in the formation process. In the first stage,
rocky cores grow by rapid planetesimal accretion. In the second stage, the core's feeding zone is depleted of solids and the atmosphere grows gradually, regulated by the KH contraction. Finally, when the atmosphere reaches the crossover mass,
gas runaway takes place and the planet gets inflated into a gas giant. The timescale of second stage is the longest among
the three and dominates whole formation process (Pollack et al. 1996).

About 20\% of Sun-like stars host super-Earths with radii of 1-4 $R_{\oplus}$ at distance 0.05-0.3 AU (Howard et al. 2010; Batalha et al. 2013; Petigura et al. 2013).
Radial velocity measurements (Weiss \& Marcy 2014) and transit timing variations (Wu \& Lithwick 2013) manifest that
 masses of these super-Earths are in the range of 2$-$20 $M_{\oplus}$.
The abundance of super-Earths presents a puzzle for the core instability theory.
This theory indicates that when a protoplanet reaches the super-Earth size, two physical processes
make the survival of super-Earths difficult, leading to a planetary ``desert" in this size range (Ida \& Lin 2004).
Super-Earths would excite density waves in PPDs and give rise to rapid type I migration.
This type of migration would cause the planet to be engulfed by its host star if the disk
inner edge touches the stellar surface.
Recent studies have sought various remedies
for type I migration (Yu et al. 2010; Fung \& Chiang 2017).
PPDs are expected to have an inner edge at the stellar magnetosphere (e.g., Long et al. 2005). For planets undergoing disk-driven migration, they are expected to pile up near this edge. They would stay either at the edge because the gas runs out, or inside the edge down to 2:1 resonance because that's where the tidal torque will taper off, or outside the edge as the standing waves generated by wave reflection off the inner edge stall planet migration (Tsang 2011).
In this paper we would focus on another threat for super-Earths.
Super Earths have low mean density, which suggests that they must
be surrounded by gas envelopes (Rogers \& Seager 2010).
Since these observed super-Earths are in the range of critical mass,
they would trigger efficient gas runaway and accumulate massive gas envelope.
They would become gas giants. As a result, super-Earths are supposed to be rare.
However, the Kepler's discovery wreck these predictions. Lee et al. (2014) has proposed
metallicity gradient inside the PPDs or late assembly of cores to
resolve the puzzle of super-Earth formation. Lee \& Chiang (2016) stressed
that the late core assembly in transitional PPDs is more consistent with
observations. In gas-poor environments, gas dynamical friction has
weakened to allow proto-cores to stir one another and merge.
In addition, this formation scenario ensures that super-Earth cores accrete mass
with a few percent envelope mass fraction (EMF).



Guillot \& Showman (2002) argued that the dissipation of kinetic energy of atmospheric wind,
driven by intense irradiation, could bury heat inside the planet.
Many studies extend this idea to explain the radius anomaly of hot Jupiters
(Youdin \& Mitchell 2010; Ginzburg \& Sari 2015; Komacek \& Youdin 2017).
These investigations focus on the late evolution after the disk dispersal.
Unfortunately, this is invalid for the early evolution of super-Earths because they are still
embedded within disks. The irradiation may not penetrate the
disk and is not able to bury heat in the exoplanets.

However, we note that
tidal interactions between the host star and planet can periodically perturb the planet
and generate mechanical forcing of the fluid motions (Zahn 1977; Goldreich \& Nicholson 1989).
Heating by tidal dissipation in primordial super-Earth envelope can inhibit the gas cooling (Ginzburg \& Sari 2017).
This mechanism requires the orbital eccentricity of super-Earths be continuously pumped.
But super-Earths may not be massive enough to clear a clean gap to excite orbital
eccentricity (Goldreich \& Sari 2003).
Another important aspect about tidal interaction is that  
tidally-forced turbulent mixing would induce heat transport inside the planets.
Recent laboratory experiment shows that turbulence could penetrate deep inside the
planet interior (Cabanes et al. 2017).
By combining laboratory measurements and high resolution simulations, Grannan et al. (2017)
confirmed the generation of bulk filling turbulence inside planet driven by tidal forcing.
Turbulent mixing plays an essential role in heat transport in strongly stratified environments (Garaud \& Kulenthirarajah 2016).
This motivates us to study the effects of turbulent diffusion on the planet's thermal evolution.
Prior studies have noticed that the turbulent mixing induced by mechanical forcing
leads to heat transport inside hot Jupiters (Youdin \& Mitchell 2010).
These tides would produce appreciable thermal feedback and may lead to interior radiative zones, enhancing
g-mode dissipations with a wide spectrum of resonances (Jermyn et al. 2017).
We find that the thermal feedback associated with the externally-forced turbulent stirring
may greatly alter the accretion history of super-Earths.

It is well known that the timescale of gas accretion is dictated by the KH timescale.
In other words, accretion is determined by the planet's ability to cool (Lee \& Chiang 2015).
In this paper, we note that 
the tidally-forced turbulent diffusion influences the heat transport inside the planet's envelope.
Thermal feedback would be induced by turbulent diffusion.
The heat transport associated with tidally-forced turbulent diffusion
would reduce the cooling luminosity
and enhance the KH timescale. 
We find that turbulent diffusion may have significant effects
on the planet accretion history\footnote{In our
calculation, turbulent diffusion coefficient $\mu_{\rm turb}\sim 10^{7} - 10^{9}$ cm$^2$ s$^{-1}$,
comparable to typical
$\mu_{\rm turb}\sim 10^{6}-10^{10}$ cm$^2$ s$^{-1}$ in solar system planets (de Pater \& Lissauer  2001).}.
Based on our calculations, we propose that tidally-forced turbulent diffusion would effectively
help super-Earths evade growing into gas giants.


This paper is structured as follows. In section 2, we provide a brief description of the accreting
planet envelope with tidally-forced turbulent diffusion.
In section 3, we compare the planet interior thermal
profile with and without turbulent diffusion, discussing the thermal feedback
induced by turbulent diffusion, especially the shift of RCBs.
In section 4, we depict the cooling luminosity variations and onset of gas runaway.
The quasi-static Kelvin-Helmholtz evolution and critical turbulent diffusivity
are discussed in Section 5. In Section 6, we discuss the mass loss
mechanisms for super-Earths and the limitation of super-Earth formation by the turbulent diffusion.
Summary and conclusions are given
in Section 7.

\section{Accreting Envelope with Tidally-Forced Turbulence}
Super-Earths are susceptible to runaway accretion (Pollack et al. 1996).
The ability to accrete is determined by the planet's power to cool (Lee \& Chiang 2015).
How super-Earths avoid rapid gas runaway
depends critically on the cooling history of planet, which is closely related to
the thermal structure of the envelope.
In the convectively stable region, the turbulent diffusion would induce heat transport within the planet.
In this section we will concentrate on the thermal feedback caused by tidally-induced turbulent diffusion.

\subsection{Thermal structure of Gaseous Envelope}
Since the planet's ability to cool depends on planets' thermal structure of the envelope,
we first study the gaseous envelope structure of planets,  i.e.,  the distribution of pressure,
temperature, and mass around a protoplanetary core with mass $M_{\rm c}$ embedded
within the protoplanetary nebular. 
The planet envelope (or, interchangeably ``atmosphere'') structure is governed by the following equations of mass
conservation, hydrostatic equilibrium, thermal gradient, and energy conservation
(Kippenhahn et al. 2012) :
\begin{equation}
\frac{d M_r}{d r} = 4 \pi \rho r^2 \ ,
\end{equation}
\begin{equation}
\frac{d P}{d r} = - \frac{G M_r}{r^2} \rho \ ,
\end{equation}
\begin{equation}
\frac{d T}{d r} = \nabla_{\rm } \frac{T}{P} \frac{d P}{d r} \ ,
\end{equation}
\be
\frac{d L}{d r} = \frac{d M_r}{d r} \left( \epsilon - T\frac{\partial s}{\partial t} \right) \ ,
\ee
where $G$  is  the gravitational constant, $P$ is the pressure, $\rho$ is the density,
$T$ is the temperature,  $L$ is the luminosity,
and $M_r$ is the mass, including the core mass
and the atmosphere mass,  enclosed inside the radius $r$,
$M_r = M_{\rm atm} + M_{\rm c}$.
The symbol ``$\nabla$" denotes the temperature gradient
inside the envelope.
The energy generation $\epsilon$ is set to zero since there is no nuclear reaction
inside the planet.
The above equations implicitly indicate
that the envelope quickly adjusts and dynamical timescale is shorter
than the accretion timescale (Rafikov 2006).
Note that the right hand side term, $-T\frac{\partial s}{\partial t}$, in the energy equation dictates the cooling process.
Replacing the local energy equation by a global energy equation would greatly reduce the
numerical tasks and we need only deal with ODEs rather than PDEs
(Piso \& Youdin 2014; Lee et al. 2014). Details will be discussed in Section 3.

The energy transport in the convective region is very efficient and
the temperature gradient is\footnote{This assumption is mainly made for simplicity
of the models, they are not necessarily correct (Stevenson 1985; Leconte \& Chabrier 2012).
We are working on including the mixing length theory (e.g Kippenhahn et al. 2012) to
better quantify the issue of super-adiabaticity.}
\be
\nabla = \nabla_{\rm ad} = \left( \frac{d \ln T}{d \ln P}\right)_{\rm ad} \ .
\ee
The convective and radiative layers of the envelope are specified by
the Schwarzschild criterion: the atmosphere is stable against
convection when $\nabla < \nabla_{\rm ad} $ and convectively
unstable when $\nabla \ge \nabla_{\rm ad} $. Since the convective energy
transport is efficient, $\nabla = \nabla_{\rm ad}$ in the convective region.
The actual temperature gradient can be expressed as
\be
\nabla_{\rm } = \min(\nabla_{\rm ad}, \nabla_{\rm rad}) \ .
\ee
In this paper, we adopt a polytropic index $\gamma=7/5$ for an ideal
diatomic gas and the adiabatic gradient $\nabla_{\rm ad} = (\gamma-1)/\gamma$.
Note that the realistic equation of state (EOS) would change the value of $\nabla_{\rm ad}$
and the effects of realistic EOS will be left for future studies.

The radiative temperature gradient
\be
\label{radTgrad}
\nabla_{\rm rad} = \frac{3 \kappa L P}{64\pi\sigma G M_r T^4} \ ,
\ee
where $\kappa$ is the opacity.  
In the upper part of the atmosphere, the exact
value of $\kappa$ is highly uncertain because the amount of dust and the dust size
distribution are not well constrained in PPDs.
Lee et al. (2014) studied both dusty and dust-free atmosphere and
found that the radiative-convective boundaries (RCBs) are determined by
H$_2$ dissociation at an almost fixed temperature $\sim$2500 K for dusty atmosphere.
They also found the for dust-free atmosphere, the radiative region keeps an almost
isothermal temperature fixed by the envelope outer surface.
Technically, the opacity laws can be written as a power law as a function
of pressure and temperature whether or not the total opacity is dominated
by dust grains. For these reasons, we adopt a power law opacity
(Rafikov 2006; Piso \& Youdin 2014; Ginzburg et al. 2016), by assuming that
\be
\kappa = \kappa_0 (P/P_0)^{\alpha} (T/T_0)^{\beta} \ .
\ee
Here we choose  $\kappa_0 = 0.001$cm$^{2}$g$^{-1}$,  which allows our
fiducial model without turbulent diffusion to possess properties of more
sophisticated super-Earth models (Lee et al. 2014).
What is important is the opacity near the RCB. In that sense,
it is important to keep in mind that the power-law indices
$\alpha$ and $\beta$ can change significantly within the envelope (and with distance from the star).
We have tried different choices of $\alpha$ and $\beta$. We find that, as long as the
parameter $\alpha$ and $\beta$ satisfy $\nabla_0 \equiv \frac{1+\alpha}{4-\beta} > \nabla_{\rm ad}$,
our results are robust and insensitive to the choices we made\footnote{
In later part of this paper, we present the results with $\alpha=1$, $\beta = 1$,
which ensures the existence of the inner convective region and outer radiative region
inside the planet gas envelope.  For details, please refer to discussions in Rafikov (2006)
and  Youdin \& Mitchell (2010).}.




Conventionally, it is believed that solid cores accrete planetesimal
and gas simultaneously (Pollack et al. 1996; Bodenheimer et al.
2000).
However, estimation shows that the termination epoch
of accretion of solids is well before the accretion of gas.
The dust coagulation timescale can be as short as
$t_{\rm coagulate} \sim 10^4$ yr especially
when the planet is close to the central host (Lee et al. 2014).
This timescale is much shorter than typical disk dispersal timescale ($\sim$ 0.5$-$10 Myr).
In addition, calculations by
Lee \& Chiang (2015) showed that planetesimal accretion does not generically prevent runaway.
As a result, it is physically valid to set the planetesimal
accretion rate to zero ($L_{\rm acc}=0$)
when we study accreting super-Earths within the disk.
In this case, the core is free to cool and contract, and it is  extremely susceptible to the gas runaway.

Note that the above differential equations are essentially identical
to the usual planet interior structure equations. The distinction is the thermal
feedback generated by tidally-forced turbulent mixing inside the stably stratified region.
More specifically, $\nabla_{\rm rad}$ is affected by the turbulent diffusion, which will
be further discussed in the next section.

\subsection{Thermal Feedback by Tidally-Forced Turbulent Mixing}
How do super-Earths evade becoming gas giants? 
In this paper, we propose a robust mechanism to avoid runaway accretion.
Due to the tidal forcing, the planet's gas envelope would be stirred and
the turbulent motion may be initiated.
Detailed analyses of these processes are rather complex
and beyond the scope of this paper (e.g., Garaud \& Kulenthirarajah 2016; Grannan et al. 2017).
In this paper, we try to constrain the turbulent diffusion
that is necessary to influentially affect the planet accretion timescale.
We find 
that even weak turbulence would affect the planet accretion history significantly.

Since the sound-crossing time is much shorter than
the time for heat to diffuse across the fluid blob, the blob conserves entropy (i.e. adiabatically) and keeps
pressure equilibrium with the ambient environments when it displaces over a radial
distance $\ell$.
The temperature difference between the blob and its surroundings is
\be
\delta T = \left(\frac{d T}{dr}\bigg|_{\rm ad} - \frac{d T}{dr} \right) \ell =  - \frac{\ell  T}{c_p} \frac{d s}{dr} \ .
\ee
The heat excess associated with these fluid blobs can be written
as $\delta q = \rho c_p \delta T$ and the corresponding turbulent heat flux is
$F_{\rm turb} = v \delta q$, where $v$ is the characteristic speed of turbulent eddies.
The entropy gradient can be put down as 
\be
\frac{d s}{d r} = \frac{ g}{T \nabla_{\rm ad} } (\nabla_{\rm ad} - \nabla) \ ,
\ee
where $g$ is the gravitational acceleration. This equation
indicates that in the stably stratified region ($\nabla < \nabla_{\rm ad}$),
the entropy gradient is positive ($ds/dr>0$). The heat flux by turbulent mixing is
then
\be
F_{\rm turb} = v\delta q = \rho c_p v \delta T = - \rho g v  \ell
 \left( 1- \frac{\nabla}{\nabla_{\rm ad}} \right) \ .
\ee
The flux is negative for stable stratification.
For a thermal
engine without external forcing, heat always flows from hot to cold regions.
However, with external mechanical forcing by tides, heat flows from cold
to hot regions becomes feasible (Youdin \& Mitchell 2010).
Note that the turbulent diffusion coefficient $\mu_{\rm turb} \equiv v \ell$\footnote{Note that $\mu_{\rm turb} = K_{zz}$, a symbol widely used in the community of planetary atmospheres.} and the corresponding luminosity  is
\be
L_{\rm turb} = 4 \pi r^2  \left[ - \rho g \mu_{\rm turb}  \left( 1- \frac{\nabla}{\nabla_{\rm ad}} \right) \right] \ .
\ee
The total luminosity is carried by two components, the radiative and the turbulent
\be
L = L_{\rm rad} + L_{\rm turb}   \  .
\ee
We note that the temperature gradient in the radiative region
can be arranged in a compact form as (see Appendix A for details),
\be
\nabla^{\rm }_{\rm rad} = \frac{1 + \eta }{1/\nabla^{(0)}_{\rm rad} + \eta/\nabla_{\rm ad}}  \  .
\ee
In the above equation,
\be
\nabla^{(0)}_{\rm rad} \equiv \frac{3 \kappa P L}{64\pi\sigma G M_r T^4} \ ,
\ee
and
\be
\eta \equiv \frac{4 \pi \mu_{\rm turb} G M_r \rho}{L} = 4 \pi \left( \frac{M_{\rm c}}{M_{\oplus}} \right) \nu_{\rm turb} \left(\frac{M_r}{M_{\rm c}} \right)  \left( \frac{\rho}{\rho_{\rm disk}} \right)   \ ,
\ee
where the superscript ``(0)" indicates the radiative temperature gradient without turbulence\footnote{This equation is actually
the same as the equation (\ref{radTgrad}) in this paper.} and  $M_{\rm c}$ is the mass of the solid core.
It can be readily shown that the following inequality holds in
radiative region  $\nabla^{(0)}_{\rm rad} < \nabla < \nabla_{\rm ad}$ (see Figure 3 for the
pseudo-adiabatic region).
Here we stress that it is the turbulent diffusion driven by external tidal forcing that makes $\nabla$
steeper than $\nabla_{\rm rad}^{(0)}$. This inequality has significant implications for
the thermal feedback induced by tidally-forced turbulent diffusion.
An interesting issue is that radiative zones would be enlarged and the cooling luminosity
would be greatly reduced. 

Here we define two dimensionless parameters
\be
\label{turb_def}
\nu_{\rm turb} \equiv \frac{\mu_{\rm turb}}{L/(GM_{\oplus}\rho_{\rm disk})} \ , \ \zeta \equiv \frac{\mu_{\rm turb}} { H_p c_s} \ .
\ee
The two parameters represent the strength of turbulence. In the definition of $\zeta$,
$H_p\equiv -d r/d\ln P$ and $c_s$ are pressure scale height and sound speed, respectively.
It is obvious that, if the turbulence in the radiative region is negligible, i.e., $\eta = 0$,
the temperature gradient recovers its usual definition,
$\nabla_{\rm rad} \rightarrow \nabla^{(0)}_{\rm rad}$.
In section 5.1, we will give a physical estimation of the parameter $\zeta$ based on our calculations. We will
see that small value of $\zeta \sim 10^{-6} - 10^{-5}$ has already appreciable effects on the
formation of super-Earths. This mechanism is robust in the sense that even weak turbulence is
adequate for it to operate.
We should keep in mind that one limitation is that the turbulence strength is parameterized,
not physically specified. This is an important issue which still remains to be addressed, i.e.,
forcing turbulence induced by tides
should be investigated in further detail (Barker 2016; Grannan et al. 2017).

\subsection{Boundary Conditions}
The density and temperature at the outer boundary of the atmosphere are given
by the nebular density and temperature. We adopt the minimum mass extrasolar nebula (MMEN) model
of Chiang \& Laughlin (2013). According to MMEN, the disk structure reads,
\be
\rho_{\rm disk}= 6\times 10^{-6} \left(\frac{a}{0.1 {\rm AU}} \right)^{-2.9} {\rm g \ cm^{-3}} \ ,
\ee
\be
T_{\rm disk} = 1000 \left( \frac{a}{0.1 {\rm AU}} \right)^{-3/7} {\rm K} \ .
\ee

The inner boundary lies at the surface of the inner core.
The core density is assumed to be $\rho_{\rm core} = 7$g cm$^{-3}$, the core mass
is 5 $M_{\oplus}$ and the core radius is $R_{\rm core}$ = 1.6 $R_{\oplus}$.
The outer boundary condition is chosen at the smaller of the Bondi radius and Hill radius,
which are
\be
R_H \approx 40 R_{\oplus} \left[ \frac{(1+{\rm EMF}) M_{\rm core}}{5 M_{\oplus}}\right]^{1/3} \left( \frac{a}{0.1 {\rm AU}}\right)  \ ,
\ee
\be
R_B \approx 90 R_{\oplus} \left[ \frac{(1+{\rm EMF}) M_{\rm core}}{5 M_{\oplus}}\right] \left( \frac{1000 {\rm K}}{T}\right)  \ ,
\ee
respectively. 


\section{Thermal Properties of Gas Envelopes}
Since the thermal cooling timescale is intimately related to the planet interior structure,
we first describe the interior structure of the gaseous envelope.
To avoid the complication induced by sandwiched convection-radiation structure
inside the planet interior (Ginzburg \& Sari 2015; Jermyn et al. 2017),
we simply consider a two-layer model, i.e.,
a convective interior and a radiative exterior (Piso \& Youdin 2014).

We adopt the assumption that the luminosity, $L$, is spatially constant, which
is valid in radiative region if the thermal relaxation timescale is shorter than thermal times in the rest
of the atmosphere. The validation of such assumption is corroborated by
Piso \& Youdin (2014) and  Lee et al. (2014).
To get thermal profiles within the envelope, a luminosity $L$
is required to obtain $\nabla_{\rm rad}$ before we numerically integrate the structure
equations.
The spatially constant $L$ is treated as an eigenvalue of the ODEs.
To get the eigenvalue numerically, we first give a guess value of $L$ and
re-iterate the integration until the mass at the core, $m(R_{\rm c})$,
matches the actual mass $M_{\rm c}$. Note that, once the luminosity is found,
the location of radiative-convective boundary (RCB) can be specified accordingly.

\subsection{Envelopes without Heat Transport by Turbulent Mixing}
\begin{figure}
\includegraphics[scale=0.75]{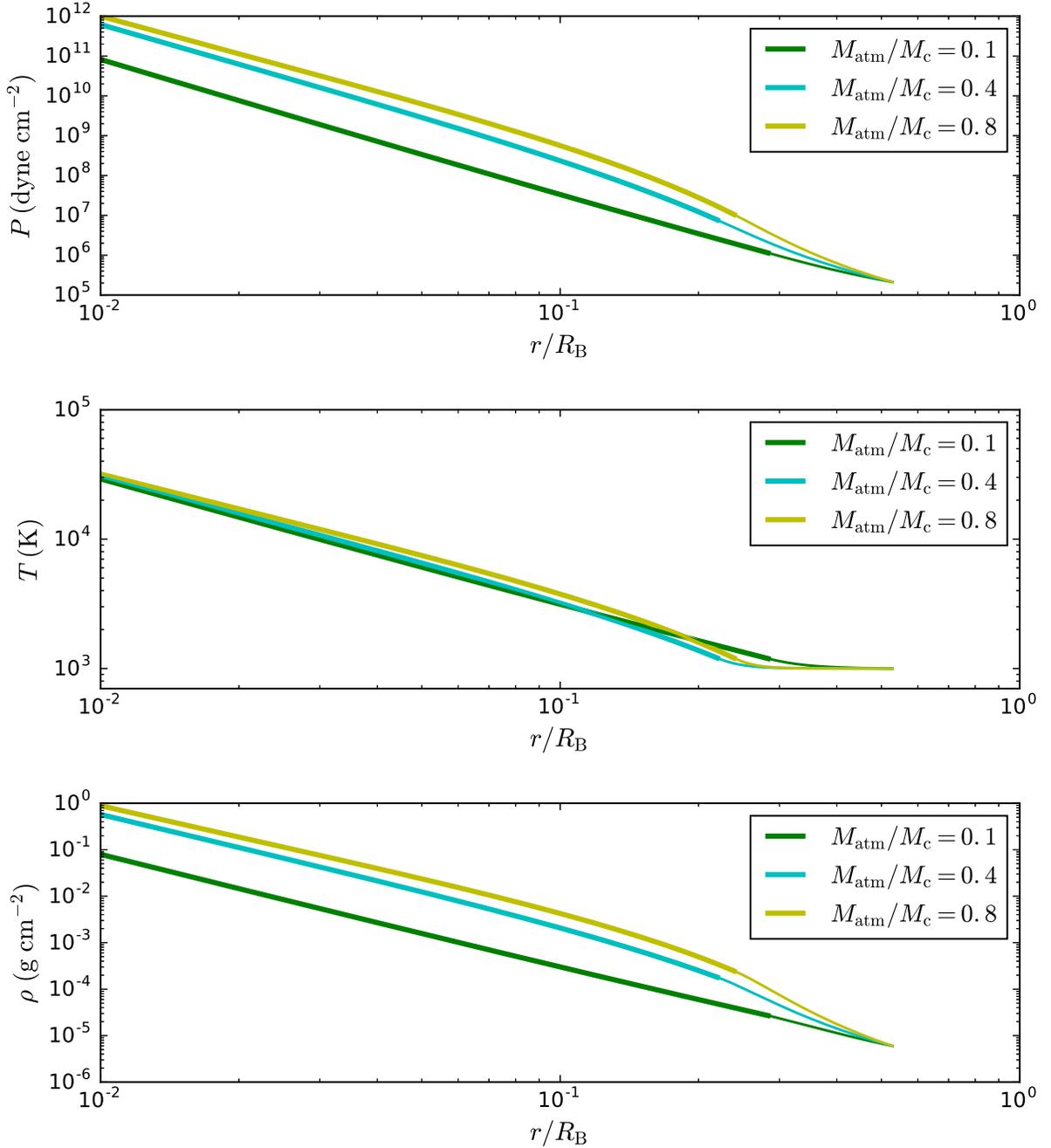}
\caption{\label{AtmProfile}
Thermal profiles around a planet core
with mass $M_{\rm c} = 5 M_{\oplus}$ at 0.1 AU.
Turbulence is not included, $\nu_{\rm turb} = 0$.
The pressure, temperature, and density are shown in the upper, middle, and lower panels,
respectively.  In each panel,
the green, cyan, yellow lines stand for $M_{\rm atm}/M_{\rm c} = $ 0.1, 0.4, 0.8,
respectively.
With the increase of envelope mass, the pressure at RCBs always increases.
However, the position of RCBs inside the planet first decreases and then increases.
This non-monotonic behavior is
due to the effects of self-gravity (Piso \& Youdin 2014).
Note in particular that no pseudo-adiabatic region appears in the envelope (cf. Figure \ref{AtmProfile_with_turb}).
}
\end{figure}

For the convenience of comparison, we first consider a fiducial model, i.e.,
an envelope without turbulence ($\nu_{\rm turb} = 0$).
In Figure \ref{AtmProfile}, we show the radial profiles of pressure, temperature, and density of
the envelope for a 5$M_{\oplus}$ core with increasing envelope mass during atmospheric growth.
The green, cyan and yellow curves denote the envelope mass fraction (EMF) = 0.1, 0.4, 0.8, respectively.
The thicker and thinner parts stand for the convective and radiative region, respectively.
The boundaries of the thicker and thinner part are the radiative-convective boundaries (RCBs).
The convective region is adiabatic.
The radiative region connects the lower entropy interior to the higher entropy exterior.
In Figure \ref{AtmProfile}, we note that the pressure in the convection zone increases with
envelope mass, but the temperature only varies slightly.
Since the entropy is $\propto \ln(T^{1/\nabla_{\rm ad}}/P)$,
it is clear that, with increasing envelope mass,
the steady-state envelopes evolve in order of decreasing entropy (Marleau \& Cumming 2014).
This is consistent with the cooling process that the envelope experiences,
which allows the atmosphere to accrete more gas.

Lee et al. (2014) found that, for dusty atmosphere, the location of RCBs lies
at an roughly fixed temperature where H$_2$ dissociates ($\sim$ 2500K).
In Figure \ref{AtmProfile}, the RCB lies at the bottom
of the outermost radiative region and the temperatures at the RCBs are no longer 2500K.
This is because we adopt a grain-free atmosphere due to efficient grain coagulation (Ormel 2014).
According to the middle panel of Figure \ref{AtmProfile},
we find that grain-free atmosphere behaves differently from grain-rich atmosphere.
The outer radiative region is nearly isothermal, which implies that $T_{\rm RCB} \sim T_{\rm out}$.
Such features have also been identified in Lee \& Chiang (2015, 2016) and Inamdar \& Schlichting (2015),
which can be readily understood 
in terms of the following relation (Rafikov 2006; Piso \& Youdin 2014)
\be
\frac{T_{\rm RCB}}{ \ T_{\rm out} }  \sim \left(1 - \frac{\nabla_{\rm ad}}{\nabla_0} \right)^{-1/(4-\beta)}  \sim 1 \  .
\ee
The term on the right hand side of this equation is around the order of unity.
This explains why the temperature at RCB, $T_{\rm RCB} \sim T_{\rm out}$.
We would stress that, the above relation is only valid for atmosphere without turbulence.
When heat transport by turbulent mixing is taken into account, the RCB is pushed inwards,
and the temperature at the RCB ($T_{\rm RCB}$) becomes higher.

At the early stage of accretion, the envelope mass is small and
the envelope can be well treated as non-self-gravitating.
In this case, simple analytic results can be derived (Rafikov 2006; Piso \& Youdin 2014).
Though the envelope we consider in this paper is self-gravitating, these analytical results
are still very instructive to understand atmospheric evolution and interpret our
numerical results.  How the position of the RCBs
varies with envelope mass can be understood with the following relations (Piso \& Youdin 2014),
\be
\label{LMrelation}
\frac{M_{\rm atm}}{M_{\rm c}} = \frac{P_{\rm RCB}}{\xi P_{\rm M}}  \ , \
\frac{P_{\rm RCB}}{ \ P_{\rm disk} } \sim \ e^{R_{\rm B}/R_{\rm RCB} } \ .
\ee
where $\xi$ is a variable on the order of unity and $P_{\rm M}$ is the characteristic pressure that is
related to the core mass (Piso \& Youdin 2014).
In the early stage of planet accretion, with the increase of envelope mass, the pressure at
RCB would increase as well.
Accordingly, the cooling luminosity would be reduced.
When the self-gravity becomes important, the above relations no longer hold.
The stronger luminosity is necessary to support the more massive envelope.
With the increase of luminosity, the RCB would be shifted outward
as shown in Figure \ref{AtmProfile} (Ginzburg \& Sari 2015).


\subsection{Envelopes with Heat Transport by Turbulent Mixing}
\begin{figure}
\includegraphics[scale=0.75]{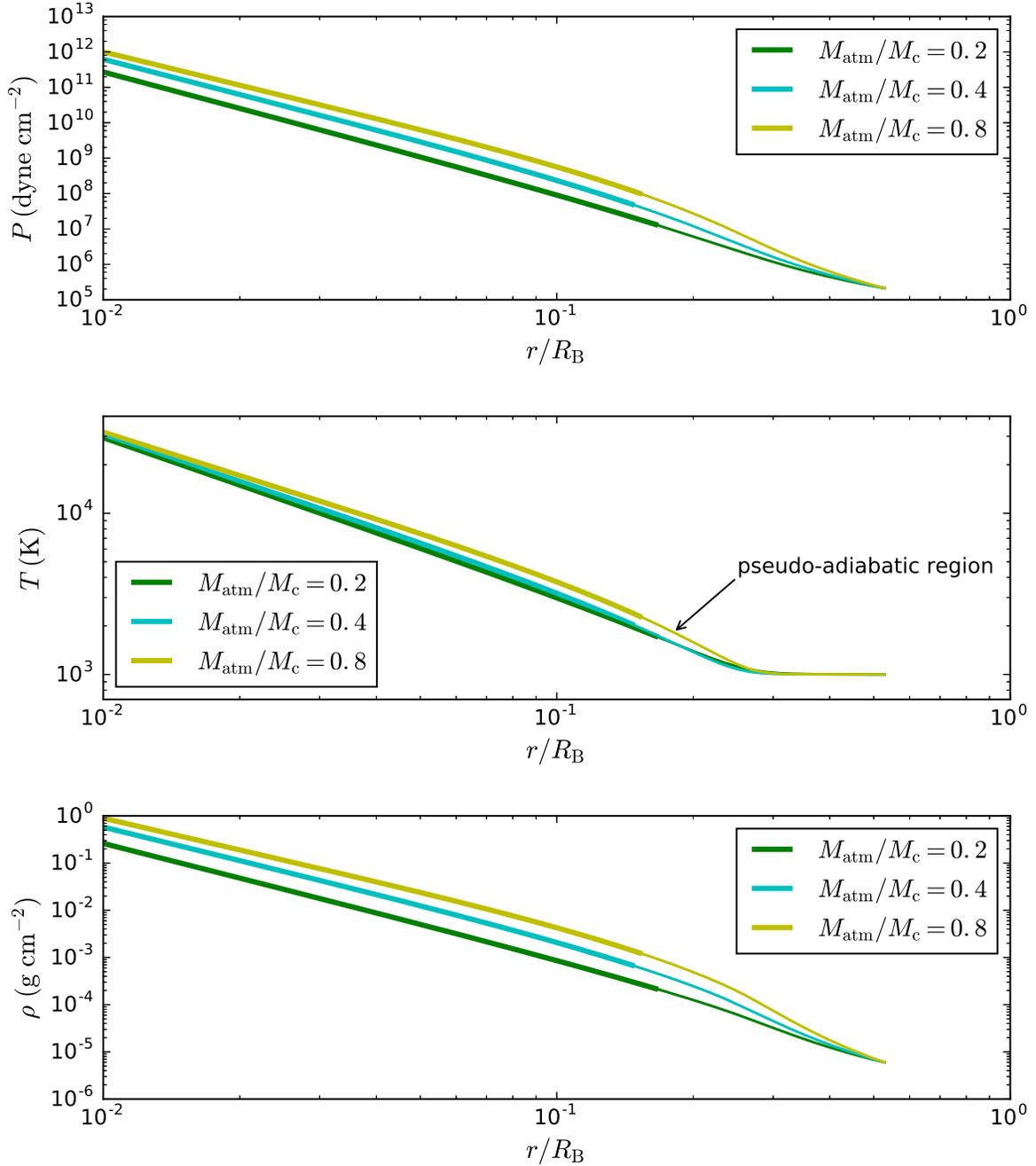}
\caption{\label{AtmProfile_with_turb}
The same as Figure \ref{AtmProfile}, but for a turbulent envelope with $\nu_{\rm turb} = 0.016$.
The pseudo-adiabatic region is most clearly visible when comparing the middle panel of Figure 1 and Figure 2.
Due to the presence of pseudo-adiabatic region, the RCBs are pushed inwards. The temperature at RCBs
becomes higher when heat transport by tidally-forced turbulent mixing is taken into account.
}
\end{figure}

\begin{figure}
\includegraphics[scale=0.65]{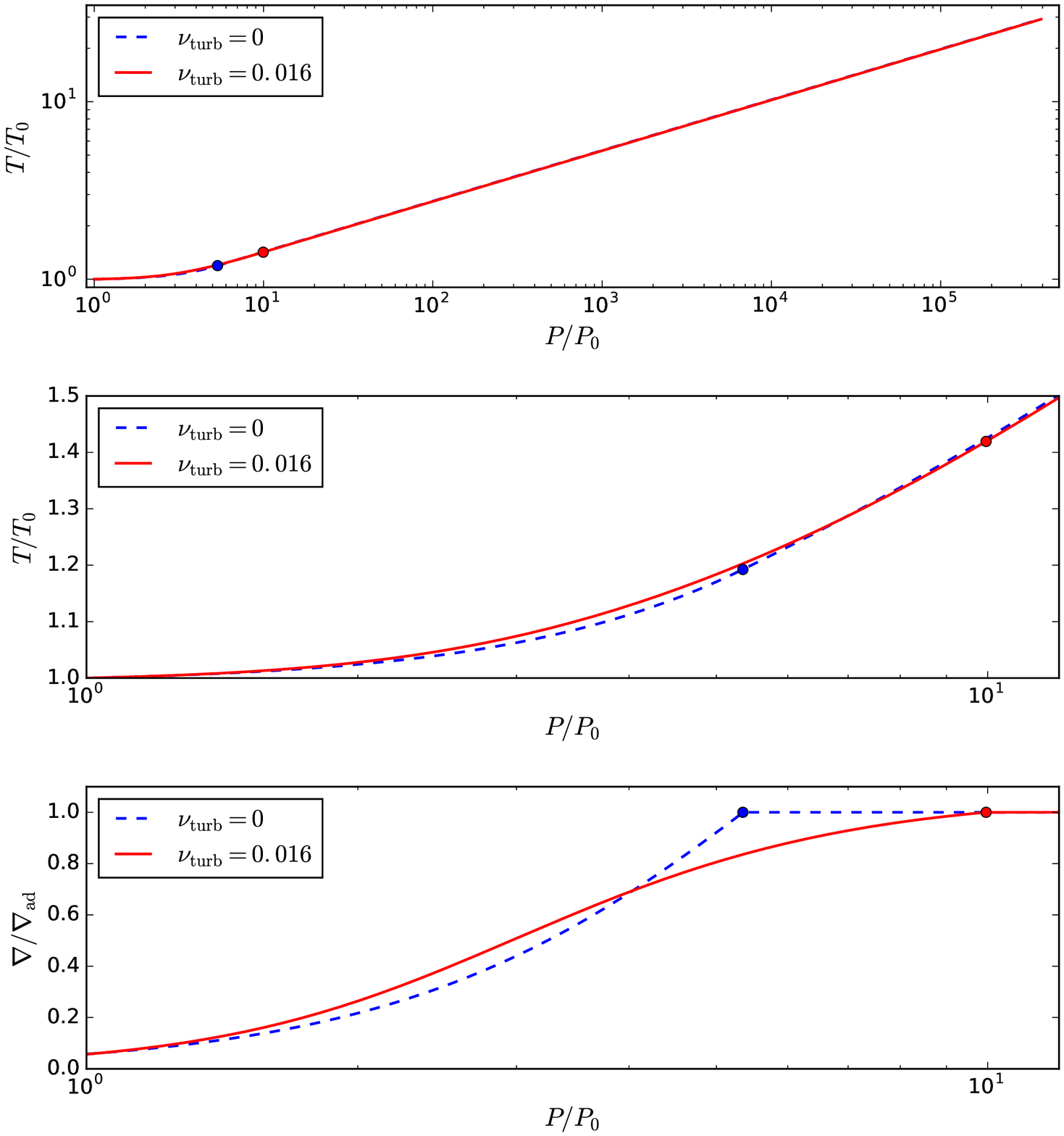}
\caption{\label{TempGradVsP}
Thermal profiles of planet envelope. The EMF  $M_{\rm atm}/M_{\rm c} =  0.1 $.
{\it Upper panel:} The blue dashed curve represents the
envelope without turbulence.
The red solid curve denotes envelope with turbulence, $\nu_{\rm turb} = 0.016$.
The RCBs are denoted as blue and red dots. 
The two temperature profiles are very similar and difficult to
distinguish.
{\it Middle panel:} To identify their differences,
we show the two profiles near the radiative-convective transition region.
The red curve shows a more gradual transition from the radiative
to adiabatic region. The region between the blue dot and red dot is the
pseudo-adiabatic region.
{\it Bottom panel:} The ratio of temperature gradient to adiabatic gradient.
The region with $\nabla/\nabla_{\rm ad} = 1$ is the convection zone.
The region with $\nabla/\nabla_{\rm ad} < 1$ is the radiative zone.
In the pseudo-adiabatic region, $\nabla$ is
very close to $\nabla_{\rm ad}$, but still smaller than $\nabla_{\rm ad}$.
The RCBs shift inwards  when heat transport by turbulent mixing
is taken into account. The RCBs penetrate deeper with stronger turbulent mixing.
 }
\end{figure}

In this section, we explore how turbulence ($\nu_{\rm turb} \neq 0$) changes the structure
of the planet envelope.
The most interesting feature is that the turbulence would push the RCBs
inwards and diminish the cooling luminosity.
In Figure \ref{AtmProfile_with_turb}, we show the planet thermal profiles for envelope
mass fraction, $M_{\rm atm}/M_{\rm c} =$ 0.2, 0.4 and 0.8.
The core mass $M_{\rm c}  = 5 M_{\oplus}$.
In Figure \ref{AtmProfile_with_turb}, we find that the difference
is that a pseudo-adiabatic region appears.
Explicitly, we point out the location of the pseudo-adiabatic region
in the middle panel of Figure \ref{AtmProfile_with_turb}.
In such regions, the temperature gradient is
very close to adiabatic gradient, but still smaller than adiabatic gradient (see Figure \ref{TempGradVsP}).

From middle panel of Figure \ref{AtmProfile}, we see that,
when the heat transport by turbulent diffusion is not included,
the RCB lies around the isothermal radiative region, $T_{\rm RCB}\sim T_{\rm out}$.
When turbulent diffusion is included, the temperature gradient would deviate
from the isothermal approximation, which is most obvious by comparing middle panels of Figure
\ref{AtmProfile} and Figure \ref{AtmProfile_with_turb}.
We can identify from Figure \ref{TempGradVsP}, that the temperature gradient near
RCBs is approaching $\nabla_{\rm ad}$ and clearly deviates from isothermal temperature
gradient. Due to this temperature gradient deviation, a pseudo-adiabatic region appears.
As a result, the temperature at RCB becomes higher and RCBs would penetrate deeper inside
the envelope.

To better understand the effects of heat transport by turbulent mixing,  we compare the profiles of planet
envelope with and without turbulence. The results are shown in Figure \ref{TempGradVsP}
as red solid and blue dashed lines, respectively.
The upper panel shows the global variation of temperature with pressure
within the envelope.
In this panel, the difference between the two cases with and without turbulence
is hardly discernible.
The middle panel shows again the variation of temperature with pressure but
focuses on the localized region around the radiative-convective transition region.
It shows that the turbulent mixing smoothes
the transition toward the adiabat.
There would appear a pseudo-adiabatic region above the
actual adiabatic region. This pseudo-adiabatic region pushes the RCB inward to higher pressure.
Turbulent mixing leads to a more gradual approach to adiabat.


The turbulent diffusion in stably stratified region provides heating, instead
of cooling so it is natural to expect that with turbulent diffusion taken into
account, the total cooling rate of envelope will decrease and KH
contraction timescale would be prolonged (see Figure \ref{LVsMatm} for details).

%

\section{Onset of Gas Runaway and Cooling Luminosity Variations}
\begin{figure}
\includegraphics[scale=0.75]{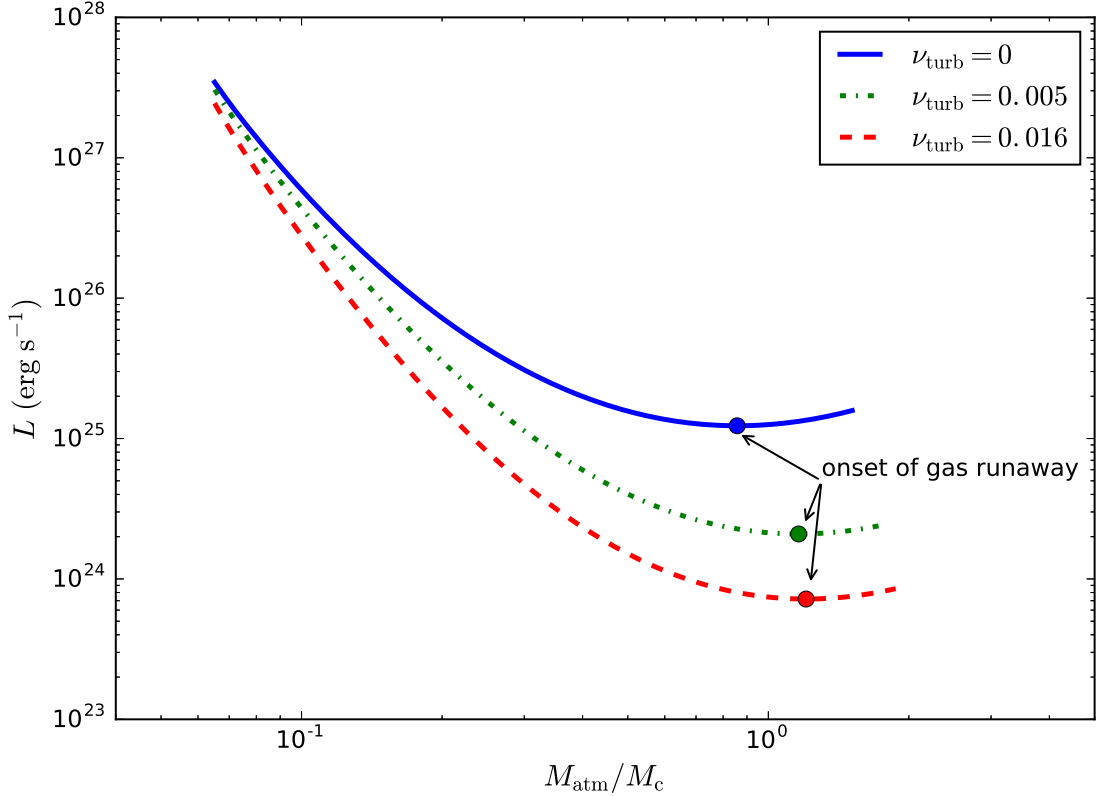}
\caption{\label{LVsMatm}
The luminosity $L$ varies non-monotonically with envelope mass.
The results for $\nu_{\rm turb} = 0,  0.005,  0.016$ are shown in
blue solid, green dot-dashed, and red dashed lines, respectively.
The luminosity minimum is reached at $M_{\rm atm}/M_{\rm c}$
= 0.86, 1.16, 1.20, respectively.
When the envelope mass is small, the increase of envelope mass
causes the luminosity to decrease. When the envelope mass is sufficiently large,
the self-gravity of gas envelope become important, and bigger
luminosity $L$ is necessary to support stronger gravity. We choose
the luminosity minimum as the epoch when the runaway accretion sets in.
We note that two important aspects of thermal evolution during the planet accretion would be
affected. With the enhanced turbulence, the cooling luminosity is reduced globally.
When the turbulence becomes stronger, the onset of gas runaway occurs
at a higher envelope mass fraction.
}
\end{figure}

Since we are interested in the planet accretion history,
it is necessary to investigate the luminosity with increasing envelope mass.
In the deep atmosphere, heat is advected by convective eddies.
Near the surface, this could be achieved by diffusion. The surface temperature
gradients would become shallower and a radiative region shows up.
The variations of luminosity with envelope mass is shown in Figure \ref{LVsMatm}.
With the accumulation of envelope mass, the luminosity reaches a minimum.
Beyond this minimum, the luminosity $L$ increases. As a result, the planet begins to cool at a very
short timescale and the envelope mass would grow super-linearly after this epoch.
Physically, it is natural to adopt the epoch when the minimum $L$ is reached as the
onset of gas runaway, $t_{\rm run}$.

On the right hand side of luminosity minimum, the luminosity-mass relation is relatively
easy to understand. At this late stage of mass growth, the self-gravity
of envelope appears to be prominent, and greater
luminosity is necessary to support stronger gravity.
However, on the left hand side of the luminosity minimum,
the mass of envelope is small and the planet is at the its early stage
of mass growth.
At this early stage (envelope's self-gravity can be ignored),
the luminosity diminishes with a thicker radiative outer layer and more massive envelope.
This reduction in cooling luminosity is intimately related to the shift of RCBs.
When the envelope self-gravity can be ignored,
the luminosity at RCB can be written as (Piso \& Youdin 2014)
\be
L_{\rm RCB} = \frac{64\pi \sigma G M_{\rm RCB} T^4_{\rm RCB}}{3 \kappa P_{\rm RCB}} \nabla_{\rm ad}
\approx \frac{L_{\rm disk} P_{\rm disk}}{P_{\rm RCB}} \ ,
\ee
where  $M_{\rm RCB}$ and $L_{\rm disk}$ reads
\be
M_{\rm RCB} = \frac{5\pi^2}{4}\rho_{\rm RCB} R_{\rm B}^{\prime}\sqrt{R_{\rm RCB}} \ , \
L_{\rm disk} \approx
\frac{64\pi\sigma G M_{\rm RCB} T_{\rm disk}^4}{3 \kappa_{\rm d} P_{\rm disk}} \nabla_{\rm ad}  \ .
\ee
The above equations can be written in terms of known properties if the envelope
mass is centrally concentrated (see, e.g., Lee \& Chiang 2015).
This central concentration is physically expected since in deeper layers where temperatures
rise above $\sim2500$K, hydrogen molecules dissociate. As energy is spent on dissociating H$_2$
molecules rather than heating up the gas, the adiabatic index drops below 4/3, to approach 1.
The upshot is that both the densities at the RCB and the radiative luminosity
can be written in terms of core properties and the temperature at the RCB.

As RCB deepens, the RCB becomes even more
optically thick so it is harder to radiate energy away; as a result, the
envelope cools more slowly.

In Figure \ref{LVsMatm}, we stress that two important aspects of thermal
evolution during the planet accretion would be
affected by turbulent mixing. The first is that it influences the luminosity.
In Figure \ref{LVsMatm}, we know that when the turbulent diffusivity ($\nu_{\rm turb}$) is enhanced, the
cooling luminosity would be reduced globally.
That is, for any particular value of envelope mass, the cooling luminosity for
an envelope with turbulence is always below that without turbulence.
When the turbulence is stronger, the luminosity becomes even smaller.
The second is that it changes the EMF at which the gas runaway occurs.
In Figure \ref{LVsMatm}, our calculations show that,
when the turbulence becomes stronger, the onset of gas runaway takes place
at higher envelope mass fraction (EMF).



\section{Quasi-Static KH Evolution and Critical Turbulent Diffusivity}

Since we ignore the accretion luminosity from the planetesimals,
the gravitational KH contraction is the only source for the cooling.
The gas accretion is regulated by the KH timescale.
Our time evolution model
can follow the envelope mass growth up to the very early epoch of
runaway growth around the crossover mass.
Fortunately, Pollack et al. (1996) found that the timescale spent in the runaway accretion stage
is orders of magnitude smaller than the KH timescale. The mass growth
timescale is  actually dominated by the KH stage. For this reason,
our model can get rather accurate estimation of mass growth timescale of an accreting planet.
In this section, we will explore how the turbulent mixing affect the KH contraction timescale.
For strong turbulent diffusion, the heat transport may even inflate the planet (Youdin \& Mitchell 2010).
We are not interested in planet inflation induced by strong turbulence in this paper. We find that even
weak turbulence can already play an essential role to delay the KH contraction.


\subsection{ Time evolution: Temporally Connecting Snapshots }
In the previous section, we have obtained snapshots of envelope
structure for different envelope masses.
To estimate the accretion timescale,
we need to connect them temporally in order of increasing mass.
The gas accretion history can be followed
by the cooling process (Piso \& Youdin 2014).
Detailed estimation shows the luminosity generated in the radiative
region can be safely ignored (Lee et al. 2014).
It is physically valid to assume the luminosity of the envelope is generated
in the convective zone and the luminosity can be treated as constant in the
outer radiative zone (Piso \& Youdin 2014).
This would greatly simplify our evolutionary calculations.
Under such circumstances, we only need to solve a set of ordinary differential equations and connect
the solutions in time.
Lee \& Chiang (2015) shows that it is physically valid to omit
planetesimal heating during the gas accretion of super-Earths.
When there is no planetesimal accretion to power the gas envelope,
the time interval between two adjacent hydrostatic snapshots is the
time it spends to cool between them.
In addition to internal energy variations, gas accretion and envelope contraction
also bring about changes to the global energy budget.
Specifically, the time interval between two steady state solutions can be written as (Piso \& Youdin 2014)
\begin{equation}\label{budget}
\Delta t = \frac{-\Delta E + \langle e\rangle\Delta M - \langle P\rangle\Delta V_{\langle M\rangle}}{\langle L\rangle} \ .
\end{equation}
Note that the symbol $\Delta$ designates the difference between
the two adjacent states and the bracket denotes the average of them.
The total energy $E$ consists of the internal energy and
the gravitational potential energy, which reads
\be
E = \int_{M_c}^{M_{}} u \ d M_r - \int_{M_c}^{M_{}} \frac{G M_r}{r} d M_r\ ,
\ee
where $u$ is the specific internal energy, $u = c_{\rm v} T$.  The second term in
equation (\ref{budget}) stands for contribution from gas accretion. The specific energy of the accreting gas
is $e = - G M_r /r + u$. The third term in equation (\ref{budget}) accounts for $P dV$ work done by the envelope
contraction.
All terms are calculated at the RCB. Note in particular that the volume difference
between two adjacent snapshots are performed at fixed mass.
We choose the fixed mass as the average of the masses at the RCB (Piso \& Youdin 2014).

\begin{figure}
\includegraphics[scale=0.65]{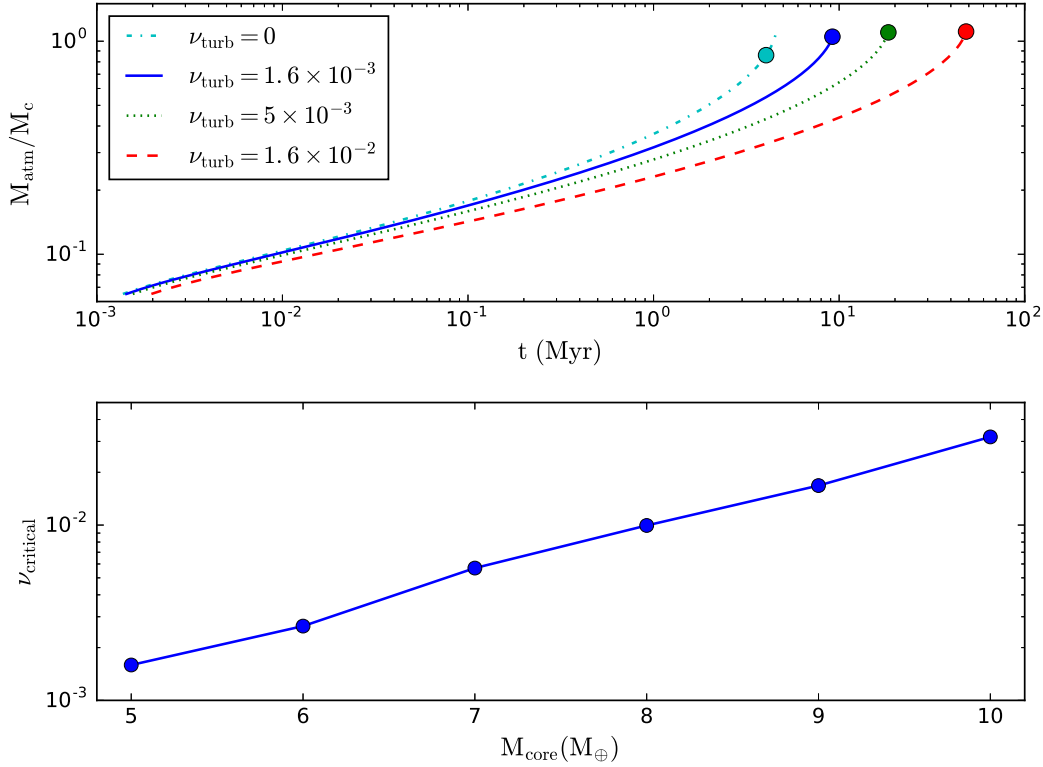}
\caption{\label{timescale}
{\it Upper panel} :
The accretion history for $\nu_{\rm turb} =$ 0, 0.0016, 0.005, and 0.016 is shown
as cyan dot-dashed, blue solid, green dotted, red dashed lines, respectively.
The initial time for the accretion is estimated as $t_0 = |E|/L$.
The slightly different starting time is due 
to the luminosity decrease by the inclusion of turbulence (see Figure \ref{LVsMatm}).
The initial EMF is around 6\%, where the planet is nearly fully convective.
Different color dots in the upper panel denote the epoch, $t_{\rm run}$, when the gas runaway takes place.
The runaway time is $t_{\rm run} =$ 4.04, 10, 18.4, and 48.3 Myrs,
respectively. The solid blue curve shows the critical solution, where $t_{\rm run} = t_{\rm disk}$.
The critical diffusivity for  $M_{\rm core} = 5 M_{\oplus}$ is
$\nu_{\rm critical}\sim 1.6\times10^{-3}$ if $t_{\rm disk}=10$ Myrs.
{\it Lower panel} : The critical $\nu_{\rm critical}$ for various core mass. For higher core
mass, the critical $\nu_{\rm critical}$ is higher.
We note that a weak turbulence with small diffusivity,
$\mu_{\rm turb} \sim 10^{7} -10^{8}$ cm$^2$ s$^{-1}$, can already enhance
the runaway timescale and delay the gas runaway.  
}
\end{figure}
In the upper panel of Figure \ref{timescale}, we shown the planet mass growth
history for different turbulent diffusivity.
In our fiducial model without turbulence, $t_{\rm run}\sim$ 4.04 Myrs.
Beyond this epoch, the gas runaway occurs.
The gas runaway is due to the fast increase of $L$ beyond $t_{\rm run}$,
which leads to a rapid cooling process on a shorter timescale.
The most intriguing feature is that
the runaway time is delayed and accretion timescale is prolonged
when heat transport by tidally-forced turbulent mixing is taken into account.
For instance, when $\nu_{\rm turb} = 0.0016, 0.005, \ 0.016$, the runaway time,
$t_{\rm run} = 10, \ 18.4, \ 48.3$ Myr, respectively.
The stronger the turbulence, the longer the gas runaway timescale.

In our calculations, we find that a small value of
$\nu_{\rm turb}$, on the order of $10^{-3}$, can already appreciably affect
the cooling timescale of super-Earths.  Since $\nu_{\rm turb}$ is dimensionless,
it is better to recover its physical value according to Equation (\ref{turb_def}).
Typically, luminosities for super-Earths are $L \sim 10^{26}$erg/s,
$M_{\oplus} = 5.97 \times 10^{27}$g, and $\rho_0 = 6\times 10^{-6}$ g cm$^{-3}$.
Then the term, $L/(GM_{\oplus}\rho_0)$, defined in Equation (\ref{turb_def})
is approximately $ \sim 4.2 \times10^{10}$ cm$^2$ s$^{-1}$.
For the dimensionless diffusivity $\nu_{\rm turb} = 0.0016$, the physical diffusivity
is approximately $\mu_{\rm turb} \sim 4.2 \times 10^{7}$ cm$^2$ s$^{-1}$.
For even larger $\nu_{\rm turb}$,
the K-H contraction timescale can be enhanced by orders of  magnitude.
According to Figure \ref{timescale}, it is evident that the turbulent diffusivity on the order
$\sim 10^7 - 10^{8}$ cm$^2$ s$^{-1}$ can already
enhance the runaway timescale by an oder of magnitude.
The pressure scale height inside the planet is $H_p \sim 10^9$ cm and the sound speed
is $c_s \sim 10^5$ cm s$^{-1}$.  
We can get a physical sense how large the turbulent diffusivity is by estimating
the dimensionless parameter $\zeta$ in Equation (\ref{turb_def}).
In our calculation, the parameter $\zeta$ is pretty small, on the order
of $10^{-7}\sim 10^{-6}$. This means that the turbulent diffusion
necessary to prolong the cooling timescale needs not to be very strong.



\subsection{Critical Turbulence Diffusivity $\nu_{\rm critical}$ and Super-Earth Formation}

A gas giant would be formed if the protoplanetary disk is still full of gas when
the planet enters the runaway accretion stage. However,
if the runway time $t_{\rm run}$ is longer than the disk lifetime $t_{\rm disk}$,
the disk gas is depleted and the planet is unable to accrete sufficient gas to become a gas giant,
then a super-Earth may be formed.
Two timescales, $t_{\rm run}$ 
and $t_{\rm disk}$, determine the ultimate destiny of the planet,
i.e., whether the planet becomes a super-Earth
or a gas giant. If $t_{\rm run} < t_{\rm disk}$, gas runaway occurs within the lifetime of
disk. The planet would get inflated by the runaway gas accretion and become a gas giant.
On the contrary, if $t_{\rm run} > t_{\rm disk}$, the disk disperses before the gas
runaway takes place. Because there is not enough gas material for the planet to accrete, the planet is
unable to become a gas giant. Usually the disk life is about $5-10$ Myr.
To be specific, we take the disk lifetime as $t_{\rm disk} = $ 10 Myrs throughout this paper.

In the upper panel of Figure \ref{timescale}, the core mass fixed at $M_{\rm core} = 5 M_{\oplus}$.
We find that there exists a critical diffusivity $\nu_{\rm critical} = 1.6\times 10^{-3}$.
When $\nu_{\rm turb}> \nu_{\rm critical}$, the K-H contraction timescale
becomes longer than the disk lifetime and the core would not be able to experience the gas runaway.
In this case, the formation of gas giants can be avoided and the formation of super-Earths becomes viable.
In the lower panel of Fig. \ref{timescale}, we show the variations of $\nu_{\rm critical}$
with $M_{\rm core}$. The critical diffusivity becomes larger when the core mass increases.
Specifically, for a 10 Earth mass core, the critical dimensionless diffusivity is approximately
$\nu_{\rm critical } = 3.2\times10^{-2}$. The actual diffusivity is about $\sim 10^{9}$ cm$^{2}$ s$^{-1}$.




\subsection{Variations of $\nu_{\rm critical}$ with Planet Location in PPDs }
Observationally, the Kepler statistics show that $\sim$20\% of Sun-like stars harbors super-Earths
at distance of 0.05-0.3 AU. By contrast, the occurrence rate
for hot Jupiters inside $\sim 0.1$ AU is only 1\%. To explain these observational features,
we consider how the turbulence affects the thermal evolution for planets
at different locations in PPDs.
The turbulent mixing considered in this paper is driven by the tides raised by the host star.
We believe that the tidally-induced turbulent mixing inside the planet
would become weaker when the planet is farther away from the host star.

\begin{figure}
\includegraphics[scale=0.7]{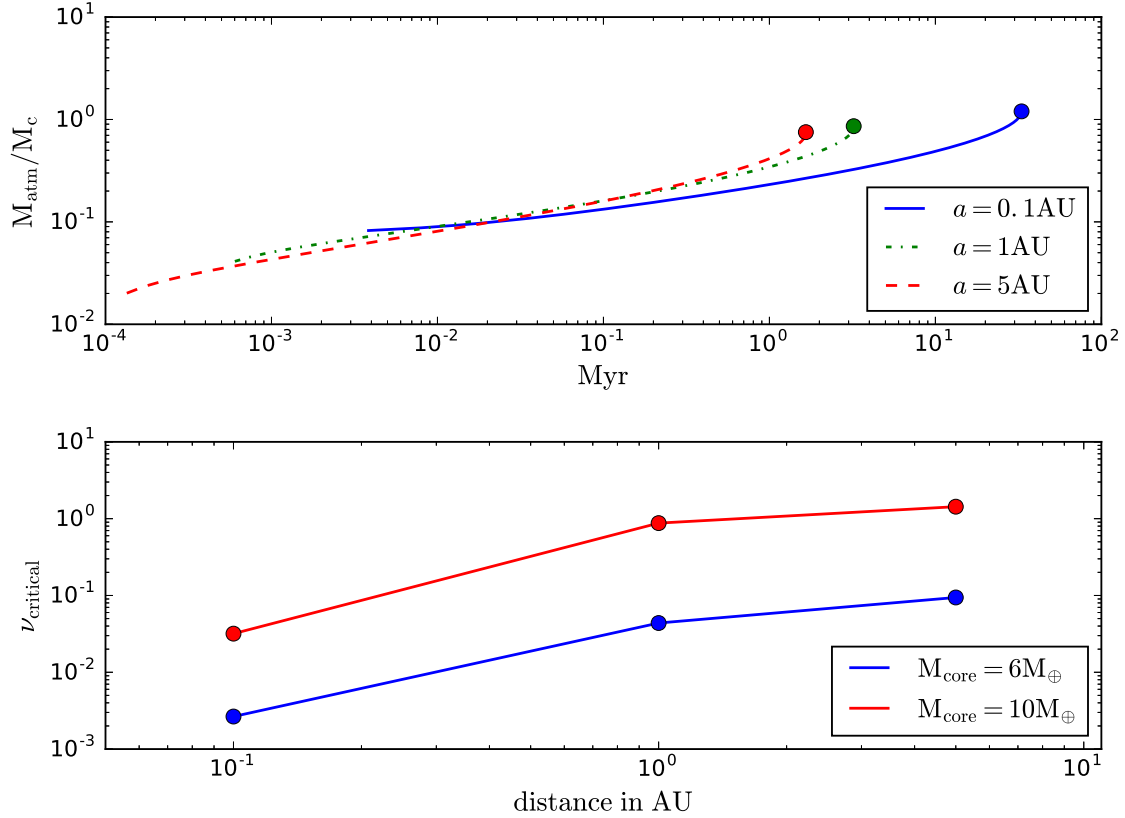}
\caption{\label{semimajor}
{\it Upper panel} : Variations of envelope mass with time.
The core mass is $M_{\rm c} = 6 M_{\oplus}$. The turbulent
diffusivity is $\nu_{\rm turb} = 0.01$. The blue solid, green dot-dashed, red dashed lines denote
mass growth history for planets at 0.1 AU, 1 AU, and 5 AU, respectively.
The critical mass ratio at the epoch of runaway decreases for more distant planet.
The runaway time for the three different cases are 33.1, 3.2, and 1.7 Myr, respectively.
It is expected that for more distant planets, larger turbulent diffusivity is required to prevent
runaway gas accretion within $t_{\rm disk} \sim$ 10 Myrs.
{\it Lower panel} : The critical diffusivity, $\nu_{\rm critical}$, for different orbital locations, required
to prevent gas runaway for disk lifetime $t_{\rm disk} \sim$ 10 Myrs.
Beyond $\nu_{\rm critical}$, the KH timescale is longer than the disk lifetime.
The formation of super-Earths becomes possible.
 }
\end{figure}

Lee et al. (2014) found that, for dusty disk, the runaway timescale is independent of
the orbital location. However, since dust can not persist in the envelope due to
coagulation and sedimentation (Ormel 2014; Mordasini 2014),  
the runaway timescale is no longer independent of the orbital location.
In the upper panel of Figure \ref{semimajor}, we show the accretion
history for planets at three different locations. The core mass is $M_{\rm c} = 6 M_{\oplus}$.
The blue solid, green dot-dashed, and red dashed curves
designate the temporal variations of envelope mass for  $a = $ 0.1AU, 1AU, and 5AU, respectively.
The turbulent diffusivity is  $\nu_{\rm turb} = 0.013$. The gas runaway occurs at
$t_{\rm run} $= 33.1, 3.2, and 1.7 Myrs.  It is clear that gas accretion
onto cores is hastened for planets that are farther away from the central star.
This behaviour can be understood from the decrease in opacity
at the RCB which makes the envelope more transparent, enhancing the rate of cooling
(Lee \& Chiang 2015; Inamdar \& Schlichting 2015).
The planet at $a = $ 1AU  and 5 AU would become a gas giant due to
runaway accretion ($t_{\rm run} < t_{\rm disk}$).
However, the planet in the inner region $a= $ 0.1 AU would become a super-Earth ($t_{\rm run} > t_{\rm disk}$).
The fact that atmospheres cool more rapidly at large distances as
dust-free worlds has been used to explain the presence of extremely puffy, low mass planets
(Inamdar \& Schlichting 2015; Lee \& Chiang 2016).

We explore the critical diffusivity, $\nu_{\rm critical}$, for planets at
different locations inside the minimum mass extrasolar nebula (MMEN).
The results are shown in the lower panel of Figure \ref{semimajor}.
It shows that the critical diffusivity increases with the semi-major axis.
When the planet is farther from the central star, $\nu_{\rm critical}$ becomes larger.
This means that the more distant planet requires stronger turbulence
to lengthen the KH timescale and avoid gas runaway.
For tidally-induced forcing, we believe that the turbulent diffusion $\nu_{\rm turb}$ is determined by
the tides inside the planet raised by host star. The tides become weaker if the planet is farther away
from the host star.
Our proposed mechanism can naturally explain the formation of close-in super-Earth,
while still ensuring the gas giant formation at larger orbital distance.
When the planet is near the host star, tidally-forced turbulent mixing is stronger and $\nu_{\rm turb}$ would be larger.
According to Figure \ref{semimajor}, the required threshold $\nu_{\rm critical}$ is smaller,
As a result, the inequality $\nu_{\rm turb} > \nu_{\rm critical}$ can be more readily
to satisfy and formation of super-Earth becomes possible. 
On the contrary, when the planet is far from the host star, $\nu_{\rm turb}$ becomes smaller as the stirring by
tides becomes weaker. The required threshold $\nu_{\rm critical}$ becomes larger.
The threshold to avoid gas runaway is more difficult to satisfy.
This indicates that, in the in-situ planet formation scenario,
it is more readily to form close-in super-Earths 
and gas giants are more prone to appear in the outer region of PPDs.
The above implication is consistent with occurrence rate inferred from observations.









\section{Mass Loss Mechanisms}
Observation shows that super-Earths possess hydrogen and helium
envelopes containing only several percent of the planet's mass.
However, we can see in Figure \ref{timescale} that the planets accrete
very massive gas envelopes.
The planet core with $\nu_{\rm turb}=0$ reaches an envelope mass fraction (EMF)
of $\sim 0.8$ at the epoch of gas runaway.
The envelope mass is considerably higher than the mass inferred from observations.
These primordial super-Earths may experience
significant mass loss during the post-formation evolution.

How super-Earths lose their mass still remains an open question.
Here we briefly mention some possible ways to lose the envelope mass.
The first possibility is that close-in planets are exposed to intense XUV (extreme UV
and X-ray) irradiation from their host stars. Photoevaporation
can significantly modify the structure of their atmosphere.
Over the timescale of $\sim 100$ Myrs, X-rays from host stars can photoevaporate
the super-Earth envelopes from initial EMF $\sim 1$ to EMF of $\sim 0.01-0.1$,
which may naturally explain the differences between
the theoretical predictions and observational facts (e.g., Murray-Clay et al. 2009;
Owen \& Wu 2013; Owen \& Wu 2017; Gaudi et al. 2017).

Giant impact is the second possible mechanism to explain the mass loss,
which is expected to be common because they are needed to provide
long-term orbital stability of planetary systems (Cossou et al. 2014).
Hydrodynamical simulations show that a single collision between similarly sized exoplanets
can easily reduce the envelope-to-core-mass ratio by a factor of two.
Super-Earths' asymptotic mass can be achieved by one or two giant impacts.
Under certain circumstances, almost 90\% of the gas envelope can be
lost during impact process (Liu et al. 2015; Inamdar \& Schlichting 2016).

Mass transfer between the close-in planet and host star via Roche lobe represent the third way to
reduce the planet mass (Valsecchi et al. 2015; Jia \& Spruit 2017; Jackson et al. 2017).
Tidal dissipation can drive orbits of these primordial super-Earths to decay toward the Roche limit.
The mass transfer is quite rapid, potentially leading to complete removal of the gaseous envelope in a few Gyr,
and leaving behind a super-Earth.
Many gaseous exoplanets in short-period orbits are on the verge or are in the process of Roche-lobe overflow (RLO).
The coupled processes of orbital evolution and RLO likely shape the observed distribution of close-in exoplanets and may even be responsible for producing some of the short-period rocky planets. But recent calculations by Dosopoulou et al. (2017) challenged this idea by claiming that,  for high eccentric planets or retrograde planets, self-accretion by
the planet would slow down the mass loss rate via Roche lobe overflow.

Super-Earth envelope mass fractions range just 1-10\% and more typically just $\sim$1\%
(see Rogers \& Seager 2010, Lopez \& Fortney 2014, Wolfgang \& Lopez 2015).
The mechanism discussed in this paper overpredicts the envelope mass fraction of super-Earths, often beyond 80\%.
Photoevaporation, even around Sun-like stars, are only effective out to ~10 days and many super-Earths lie beyond this (see, e.g., Figure 8 of Owen \& Wu 2013). Removal of $>$90\% of the envelope by giant impact requires impact velocity that exceeds the escape velocity (see, e.g., Figure 3 of Inamdar \& Schlichting 2016). Finally, Roche lobe overflow only works within ~2 stellar radii where the Roche radius is.
Lee \& Chiang (2016) proposed that the late-time formation of cores ensures that super-Earth
cores accrete a few percent envelope mass fraction, in agreement with the observations.
There is a clear difference in the expected final envelope mass fraction
between their work and ours.

Very recent works have revealed that planetary envelopes embedded within PPDs
may not be in hydrostatic balance, which slows down envelope growth. It
is possible for a steady state gas flow enters
through the poles and exits in the disc mid-plane (Lambrechts \& Lega 2017).
In the presence of a magnetic field and weakly ionizing winds,
ohmic energy is dissipated more readily for lower-mass planets.
Ohmic dissipation would make super-Earths more vulnerable to atmospheric evaporation (Pu \& Valencia 2017).
These findings may offer new explanations for the typical low-mass envelopes around the cores of Super-Earths.
In addition, we also note that the turbulent
diffusion mechanism may be still operating in the late core assembly scenario.
In the late core assembly scenario without turbulent diffusion, the asymptotic EMF is about 3-5\% (Lee \& Chiang 2016).
When turbulent diffusion is taken into account, the EMF can be further reduced to 1\%.



\section{Summary and Conclusion}

In this paper, we propose a new mechanism to avoid gas runaway for planet cores
within the lifetime of disks.
The mechanism proposed in this paper is not subject the $\kappa$ or $\mu$
catastrophe (Lee \& Chiang 2015). Tidal heating (Ginzburg \& Sari 2017) requires
orbital eccentricity be continuously pumped up during super-Earth formation.
Our mechanism does not depend on the orbital eccentricity of super-Earth.
Incorporating this model into a population synthesis model may better constrain our
understanding of the exoplanet formation (Ida  \&Lin 2004; Jin \& Mordasini 2017).

We have explored the effects of heat transport induced by tidal stirring on the thermal
structure of stably stratified, radiative layers of super-Earths,
focusing on their influences on the KH timescale.
When we take turbulent stirring into account,
pseudo-adiabatic regions would show up within the radiative zone.
This may push the RCBs inwards.
The temperature, pressure at RCBs becomes higher and the cooling luminosity would be reduced.
As a result, the KH timescale would be enhanced.
We find that
there exist a critical turbulent diffusivity $\nu_{\rm critical}$. When
$\nu_{\rm turb} > \nu_{\rm critical}$, the runaway time is greater than
the disk lifetime ($t_{\rm run}  > t_{\rm disk}$). Under such circumstances,
the onset of the planet gas runaway lags behind the disk gas depletion.
Since the planet has not enough gas to accumulate, it can no longer grow
into a gas giant and become a super-Earth instead. In addition, we also investigate
the variations of $\nu_{\rm critical}$ with planet's semi-major axis in MMEN.
Our calculations show that the condition for turbulence-induced formation of super-Earths
is more readily satisfied in the inner disk region, but is harder to satisfy in the outer
disk region. The occurrence rate of super-Earths and gas giant is consistent our calculations.

The extent of radiative region has important implication
for the tidal dissipations inside the planet.
The turbulence pushes the RCBs inwards and produces enlarged radiative zones.
Since the internal gravity waves can
propagate inside the radiative zone, the variations of this resonant cavity
would significantly influence the dissipation of internal gravity waves.
This would greatly influence the propagation and dissipation of internal
gravity waves inside the radiative zone (Jermyn et al. 2017).
Another effect is that the transition between convective zone to radiative zone is smoothed.
The radiative zone is thickened and
this bears important implications for the internal gravity wave
excitation and propagation (Lecoanet \& Quataert 2013).
This would have appreciable effects on the thermal tides inside the planet.
These issues will be addressed in a further study.

A limitation of this work is that the turbulence strength is not specified from first principle.
As a compromise, we parameterize the turbulence diffusion as a free parameter. We try
to constrain the turbulence strength in terms of the planet thermal evolution. Interestingly,
we find the turbulence in the radiative region  have substantial effects on the planet
accretion history.
How turbulence is initiated during the planet formation and how strong the turbulent diffusion
is involve very complicated physical processes, which are worth further investigations.

Realistic opacities and EOS have influential effects on
the planetary thermal structure and the core accretion process (e.g. Stevenson 1982;
Ikoma et al. 2000; Rafikov 2006),
especially for timescale of the KH timescale (Lee et al. 2014; Piso \& Youdin 2014).
Our simple prescription of opacity needs to be improved.
Guillot et al. (1994)
showed that an convective layer lies between two adjacent radiative
region due to the opacity window near $\sim$ 2000K.
A relevant caveat is the existence of radiative zones sandwiched inside convective interior.
Such radiative windows are ignored in our two-layer models.
It would be interesting
to consider how a downward turbulent heat flux would interact with such
a sandwiched region. In summary, how super-Earth envelope cooling
history responds to more realistic opacities and EOS needs to be further investigated.
Calculation with realistic EOS and opacity are underway and will be reported elsewhere.


We have found that the epoch of runaway accretion can be effectively
delayed by the turbulent diffusion within the stably stratified region. But we should be cautious that
the envelope mass fraction predicted by this mechanism is not fully consistent with observations.
The envelope mass fraction for planet embedded within the gas-rich MMEN is greater than 80\%, much higher
than the typical super-Earth envelope.
It is difficult for the turbulent diffusion alone to make the envelope mass fraction be consistent with observations.
Additional physical process, such as giant impact, photo-evaporation, Roche-lobe overflow may be operating to
reduce the envelope mass fraction during
the formation of super-Earth.  But these mass loss processes either operate on distances shorter
than most super-Earths or are applicable under certain circumstances. A promising mechanism for super-Earth formation
is the late core assembly within the transitional PPDs. In this scenario, with the reduction of the
PPD mass density, the envelope mass fraction can be as low as 3-5\% (Lee \& Chiang 2016). We note that the turbulent diffusion
may be still working in the late core assembly scenario. How turbulent diffusion affect the envelope
mass fraction within transitional PPDs is an interesting issue worth further investigation.

\acknowledgments
We thank the anonymous referee for the thoughtful comments that greatly improve this paper.
Discussions about heat transport inside planet interior
with Yanqin Wu and Re'em Sari are highly appreciated.
This work has been supported by National Natural
Science Foundation of China (Grants 11373064,  11521303, 11733010),
Yunnan Natural Science Foundation (Grant 2014HB048)
and Yunnan Province (2017HC018).


\appendix

\section{Temperature Gradient  with Turbulence}
When the turbulent flux is taken into account, the luminosity can be written as
\be
L = 4 \pi r^2 \left( - \frac{16\sigma T^3}{3 \kappa \rho} \frac{d T}{d r}\right)
+ 4 \pi r^2 \left[ - \mu_{\rm turb} \rho g \left( 1- \frac{\nabla}{\nabla_{\rm ad}} \right) \right]  \ ,
\ee
Note that in the above equation $g = GM_r/r^2$ and $\nabla = \frac{d \ln T}{d \ln P}$. This equation
can be expressed as
%
%
%
%
%
\be
\frac{d T}{d r} = \frac{L + 4 \pi r^2 \mu_{\rm turb} \rho \frac{G M_r}{r^2}}{\frac{16\sigma T^3}{3 \kappa \rho}
+ \mu_{\rm turb}  \frac{P}{\nabla_{\rm ad} T}} \left( - \frac{1}{4 \pi r^2} \right) \ ,
\ee
If we take the pressure $P$ as the independent variables, the temperature gradient becomes
%
%
\be
\nabla = \frac{P}{T}\frac{d T}{d P} =  \frac{1 + \frac{4 \pi \mu_{\rm turb} GM_r  \rho}{L} }{\frac{64\pi GM_r \sigma T^4}{3 \kappa P L} + \frac{4\pi  \mu_{\rm turb} G M_r\rho}{L}  \frac{1}{\nabla_{\rm ad}}} = \frac{1 + \eta }{1/\nabla^{(0)}_{\rm rad}+   {\eta}/{\nabla_{\rm ad}}} \ ,
\ee
where $\eta  = {4 \pi \mu_{\rm turb} GM_r  \rho}/{L}$.


\begin{thebibliography}{99}

    \bibitem[\protect\citeauthoryear{Arras \& Socrates}{2010}]{AS10}
        Arras, P. \& Socrates, A., 2010, ApJ, 714, 1


    \bibitem[\protect\citeauthoryear{Batalha}{2013}]{Batalha13}
        Batalha, N. M., Rowe, J. F., Bryson, S. T., et al. 2013, ApJS, 204, 24



    \bibitem[\protect\citeauthoryear{Bodenheimer \& Pollack}{1986}]{BP86}
        Bodenheimer, P., \& Pollack, J. B., 1986, Icar, 67, 391

    \bibitem[\protect\citeauthoryear{Bodenheimer et al.}{2000}]{BHL2000}
        Bodenheimer, P., Hubickyj, O., \& Lissauer, J. J., 2000, Icar, 143, 2

    \bibitem[\protect\citeauthoryear{Cabanes et al.}{2017}]{Cabanes17}
        Cabanes, S., Aurnou, J., Favier, B., Le Bars, M., 2017, Nature Physics, 13, 387


    \bibitem[\protect\citeauthoryear{Chiang \& Laughlin}{2013}]{CL13}
        Chiang, E., \& Laughlin, G., S, 2013, MNRAS, 431, 3444

    \bibitem[\protect\citeauthoryear{Cossou et al.}{2014}]{Cossou14}
        Cossou, C., Raymond, S. N., Hersant, F., Pierens, A., 2014, A\&A, 569, 56

    \bibitem[\protect\citeauthoryear{de Pater \& Lissauer}{2001}]{Pater01}
        de Pater, I., \& Lissauer, J. J. 2001, Planetary Sciences (Cambridge: Cambridge
Univ. Press)

    \bibitem[\protect\citeauthoryear{Dosopoulou et al}{2017}]{Doso17}
        Dosopoulou, F., Naoz, S., Kalogera, V., 2017, ApJ, 844, 12

    \bibitem[\protect\citeauthoryear{Garaud16}{2016}]{Garaud16}
        Garaud, P., \& Kulenthirarajah, L. T., 2016, ApJ, 821, 49

    \bibitem[\protect\citeauthoryear{Gaudi et al.}{2016}]{Gaudi17}
        Gaudi, B. S., Stassun, K. G., Collins, K. A.,  Beatty, T. G.,  et al., 2017, Nature, 546, 514

    \bibitem[\protect\citeauthoryear{Ginzburg \& Sari}{2015}]{Ginz15}
        Ginzburg, S., Sari R., 2015, ApJ, 803, 111

    \bibitem[\protect\citeauthoryear{Ginzburg et al.}{2016}]{Ginz16}
        Ginzburg, S., Schlichting, H. E., \& Sari R., 2016, ApJ, 825, 29

    \bibitem[\protect\citeauthoryear{Ginzburg et al.}{2017}]{Ginz17}
        Ginzburg, S., \& Sari R., 2017, MNRAS, 464, 3937


    \bibitem[\protect\citeauthoryear{Goldreich \& Nicholson}{1989}]{GN89}
        Goldreich, P., \& Nicholson, P. D., 1989, ApJ, 342, 1079

    \bibitem[\protect\citeauthoryear{Foglizzo et al.}{2007}]{Fog07}
        Grannan, A. M.,  Favier, B., Le Bars M., \& Aurnou, J. M., 2017, Geophys. J. Int., 208, 1690

    \bibitem[\protect\citeauthoryear{Guillot \& Showman}{2002}]{GS02}
        Guillot, T., \& Showman, A. P., 2002, A\&A, 385, 156

    \bibitem[\protect\citeauthoryear{Harris}{1978}]{Harris78}
        Harris, A. W., 1978, LPSCI, 9, 459

    \bibitem[\protect\citeauthoryear{Howard et al.}{2010}]{Howard10}
        Howard, A W., Marcy, G W., Johnson, J. A., et al., 2010, Sci, 330, 653

    \bibitem[\protect\citeauthoryear{Ida \& Lin}{2004}]{IL04}
        Ida, \& Lin, D. N. C., 2004, ApJ, 604, 388

    \bibitem[\protect\citeauthoryear{Inamdar \& Schlichting}{2015}]{IS15}
        Inamdar, N. K., \& Schlichting, H. E., 2015, MNRAS, 448, 1751

    \bibitem[\protect\citeauthoryear{Ikoma et al.}{2000}]{Ikoma00}
        Ikoma, M. Nakazawa, K., \& Emori, H. 2000, ApJ, 537, 1013

    \bibitem[\protect\citeauthoryear{Jackson et al.}{2017}]{Jackson17}
        Jackson, B., Arras, P., Penev, K., Peacock, S., Marchant, P., 2017, ApJ, 835, 145

    \bibitem[\protect\citeauthoryear{Jermyn}{2017}]{Jermyn17}
        Jermyn, A. J., Tout, C. A., \& Ogilvie, G. I. 2017, MNRAS, 469, 1768

    \bibitem[\protect\citeauthoryear{Jia \& Spruit}{2017}]{Jia17}
        Jia S., Spruit, H. C., 2017, MNRAS, 465, 149

    \bibitem[\protect\citeauthoryear{Komacek}{2017}]{Komacek17}
        Komacek, T. D.,  Youdin, A. N., 2017, arXiv:1706.07605v1

    \bibitem[\protect\citeauthoryear{Lecoanet \& Quataert}{2013}]{LQ13}
        Lecoanet, D., \& Quataert, E., 2013, ApJ, 797, 95

    \bibitem[\protect\citeauthoryear{Lee et al. }{2014}]{LCO14}
        Lee E. J., Chiang, E. \& Ormel, C., 2014, ApJ, 797, 95

    \bibitem[\protect\citeauthoryear{Lee \& Chiang}{2015}]{LC15}
        Lee E. J., \& Chiang, E., 2015, ApJ, 811, 41

    \bibitem[\protect\citeauthoryear{Lee \& Chiang}{2016}]{LC16}
        Lee E. J., \& Chiang, E., 2016, ApJ, 817, 90

    \bibitem[\protect\citeauthoryear{Liu et al. }{2015}]{Liu15}
        Liu. S. F., Hori , Y., Lin, D. N. C., \& Asphaug, E., 2015, 812, 164

    \bibitem[\protect\citeauthoryear{Long et al. }{2005}]{Long05}
        Long, M., Romanova, M. M., \& Lovelace, R. V. E.,2005, ApJ, 634, 1214	
    \bibitem[\protect\citeauthoryear{Lopez \& Fortney }{2014}]{LF14}
        Lopez, E. D. \& Fortney, J. J., 2014, ApJ, 792, 1


    \bibitem[\protect\citeauthoryear{Marleau \& Cumming}{2014}]{Marleau14}
        Marleau, G. D., Cumming, A, 2014, MNRAS, 437, 1378

    \bibitem[\protect\citeauthoryear{Mizuno}{1980}]{Miz80}
        Mizuno, H., 1980, PThPh, 64, 544

    \bibitem[\protect\citeauthoryear{Mizuno et al.}{1978}]{Miz78}
        Mizuno, H., Nakazawa, K., \& Hayashi, C., 1978, PThPh, 60, 699


    \bibitem[\protect\citeauthoryear{Mordasini et al.}{2014}]{Mordasini14}
        	Mordasini, C., Klahr, H., Alibert, Y., Miller, N., Henning, T., 2014, A\&A, 566, 141

    \bibitem[\protect\citeauthoryear{Murray-Clay et al.}{2009}]{MCM09}
	Murray-Clay, R. A., Chiang, E. I., \& Murray, N., 2009, ApJ, 693, 23

    \bibitem[\protect\citeauthoryear{Ormel}{2014}]{Ormel14}
        Ormel, C. W., 2014, ApJL, 789, L18

    \bibitem[\protect\citeauthoryear{Owen \& Wu}{2013}]{OW13}
        Owen, J. E., \& Wu, Y., 2013, ApJ, 775, 105

    \bibitem[\protect\citeauthoryear{Owen \& Wu}{2017}]{OW17}
        Owen, J. E., \& Wu, Y., 2017, arXiv:170510810


    \bibitem[\protect\citeauthoryear{Owen \& Wu}{2017}]{OW17}
        Perri, F., \& Cameron, 1974, Icar, 22, 416

    \bibitem[\protect\citeauthoryear{Piso \& Youdin}{2014}]{PY14}
        Piso, A. M., \& Youdin, A. N., 2014, ApJ, 786, 21

     \bibitem[\protect\citeauthoryear{Petigura et al.}{2013}]{Petgura13}
        Petigura, E. A., Marcy, G. W., \& Howard, A. W., 2013, ApJ, 770, 69

    \bibitem[\protect\citeauthoryear{Pollack et al.}{1996}]{P96}
        Pollack, J. B., Hubickyj, O., Bodenheimer, P., \& Lissauer, J., 1996, Icarus, 124, 62



    \bibitem[\protect\citeauthoryear{Rafikov}{2006}]{Rafikov06}
        Rafikov, R. R., 2006, ApJ, 648, 666

    \bibitem[\protect\citeauthoryear{Rogers \& Seager}{2010}]{RS10}
        Rogers, L. A., \& Seager S., 2010, ApJ, 716, 1208 

    \bibitem[\protect\citeauthoryear{Stevenson}{1982}]{Stevenson82}
        Stevenson, D. J., 1982, P\&SS, 30, 755

    \bibitem[\protect\citeauthoryear{Tsang}{2011}]{Tsang11}
        Tsang, D., 2011, ApJ, 741, 109


    \bibitem[\protect\citeauthoryear{Valsecchi et al.}{2015}]{Valsecchi15}
        Valsecchi, F., Rappaport, S., Rasio, F. A., Marchant, P., Rogers, L. A., 2015, ApJ, 813, 101

    \bibitem[\protect\citeauthoryear{Weiss \& Marcy}{2014}]{WM14}
        Weiss, L. M., \& Marcy, G. W.,  2014, ApJL, 783, L6

    \bibitem[\protect\citeauthoryear{Wolfgang \& Lopez}{2015}]{WL15}
        Wolfgan, A. \& Lopez, E., 2015, ApJ, 806, 183

    \bibitem[\protect\citeauthoryear{Wu \& Lithwick}{2013}]{WL13}
        Wu, Y., \& Lithwick Y., 2013, ApJ, 772, 74

    \bibitem[\protect\citeauthoryear{Youdin \& Mitchell}{2010}]{YM10}
        Youdin, A. N., \& Mitchell, J. L., 2010, ApJ, 721, 1113

     \bibitem[\protect\citeauthoryear{Yu et al.}{2011}]{Yu11}
        Yu, C. Li, H., Li, S. T., Lin, D. N. C., \& Lubow, S., 2011, ApJ, 712, 198

     \bibitem[\protect\citeauthoryear{Zahn}{1977}]{Zahn77}
        Zahn, J. P., 1977, A\&A, 57, 383

\end{thebibliography}
\end{document}